\documentclass[showpacs,amsmath,amssymb, prl,twocolumn]{revtex4}
\usepackage{graphicx}
\usepackage{dcolumn}
\usepackage{bm}

\begin{document}

\title{Minimal model of point contact Andreev reflection spectroscopy of multiband superconductors}
\author{F. Romeo$^{1}$ and R. Citro$^{1,2}$}
\affiliation{$^{1}$Dipartimento di Fisica "E.R. Caianiello", Universit\`a  di Salerno, I-84084 Fisciano (SA), Italy\\
$^{2}$CNR-SPIN Salerno, I-84084 Fisciano (SA), Italy}

\date{\today}
\begin{abstract}
We formulate a minimal model of point contact Andreev reflection spectroscopy of a normal-metal/multiband superconductor interface. The theory generalizes the Blonder-Tinkham-Klapwijk (BTK) formulation to a multiband superconductor and it is based on the quantum waveguides theory. The proposed approach allows an analytic evaluation of the Andreev and normal reflection coefficients and thus is suitable for a data fitting of point contact experiments. The obtained differential conductance curves present distinctive features similar to the ones measured in the experiments on multiband systems, like the iron-based pnictides and the MgB$_{2}$.
\end{abstract}


\maketitle
\textit{Introduction}. The superconducting matter, discovered more than 100 years ago, gives the first known example of macroscopic quantum phase coherence. Bardeen, Cooper and Schriffer (BCS) provided an unifying framework able to explain this state of matter in conventional systems, where a phonon-mediated isotropic gap (\textit{s}-wave pairing) in the quasiparticle spectrum is open \cite{bcs}. The advent of high-$T_c$ superconductivity, in the 80's, has required the inclusion of the \textit{d}-wave symmetry of the order parameter within the BCS formulation \cite{tsuei2000}, despite the identification of the \textit{glue} of the condensed state remains an open issue.

In 2001, the discovery of the superconductivity in MgB$_{2}$ \cite{mgb2} has led to the concept that two in phase superconducting gaps can coexist ($s_{++}$ pairing) \cite{giubileo2001}, as theoretically suggested few years later the BCS formulation \cite{multiband-superc1959}. Recently, the observation of superconductivity in Fe-based compounds has
renewed the interest towards multiband superconductivity. The pairing symmetry of this new class of superconductors is still under debate, even though the experimental evidences seem to be favorable to the $s_{\pm}$ pairing, implying that the electron-like and the hole-like band both develop an \textit{s}-wave superconducting state with order parameters of opposite sign \cite{s_pm_symm}.

Point contact Andreev reflection spectroscopy (PCAR) represents a powerful method to probe the order parameter symmetry which characterizes the superconducting state. As done for other superconducting systems \cite{soulen98,gonnelli_mgb2_doped,greene2008_heavy-fermions-super,kashiwaya94,Piano2006}, the PCAR method has also been applied to the Fe-based superconductors \cite{chen2008,cohen09,millo08,daghero2012,greene-iron-pcar,fetese-multigap,Naidyuk2014}. The conclusions of these studies, however, appear quite confusing due to the absence of a simple theoretical framework able to capture the great variety of features associated to the Andreev reflection spectra of a normal-metal/multiband-superconductor interface (N/$s_{\pm/++}$). Indeed, despite several theoretical efforts \cite{golubovprl2009,araujo2009,linder2009,Ghaemi09} have been made, a simple theory of PCAR experiments in multiband systems is still lacking.\\
\begin{figure}[!t]
\includegraphics[clip,scale=0.48]{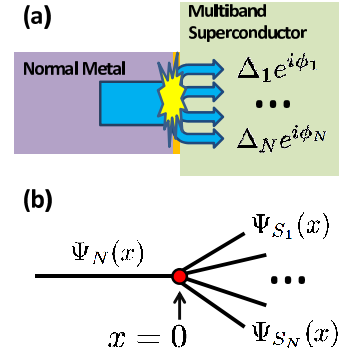}\\
\caption{(Color online) Upper panel (a): Schematic representation of a normal-metal/multiband-superconductor interface. Lower panel (b): Equivalent quantum waveguide model of the physical situation described in (a). $\Psi_{N}(x)$,  $\Psi_{S_{1}}(x)$ and $\Psi_{S_{N}}(x)$ represent the wave functions of the normal branch and of the superconducting branches, respectively.}
\label{fig:device}
\end{figure}
In this Letter we present the simplest extension of the BTK theory \cite{btk82} useful in describing the differential conductance of a N/$s_{\pm/++}$ system. Generalizing the quantum waveguides theory to include particle-hole scattering processes, analytic expressions of the Andreev and normal reflection coefficients are derived, the latter being of direct use in the data analysis. We demonstrate that, using a limited number of parameters, several distinctive features of PCAR experiments on both MgB$_{2}$ and Fe-based superconductors can be easily explained.\\
\textit{Theoretical description}. We model the ballistic interface \cite{note-ballistic} between a normal tip and a multiband superconductor by using the quantum waveguides theory \cite{xia_QWG,araujo2009} (see Fig.~\ref{fig:device}). In this framework, the physical system is represented by a network formed by interconnected one-dimensional branches. Each superconducting branch $S_{i}$, representing the $i-$th band, is described by a local Bogoliubov-de Gennes Hamiltonian. The eigenvalues problem for the wave function $\Psi_{S_{i}}$ is thus written as
\begin{equation}
\label{eq:hamiltonian}
\left[
  \begin{array}{cc}
    \hat{H}_{i}(x) & \Delta_{i} \\
    \Delta^{\ast}_{i} & -\hat{H}_{i}^{\ast}(x) \\
  \end{array}
\right]\Psi_{S_{i}}(x)=E\Psi_{S_{i}}(x),
\end{equation}
where $\Delta_{i}=|\Delta_{i}|e^{i\phi_{i}}$ represents the gap of the $i-$th band, while the single particle Hamiltonian takes the form:
\begin{equation}
\hat{H}_{i}(x)=\Bigl[-\frac{\hbar^2\partial_{x}^2}{2m_{S_{i}}}-E_F+V^{(i)}_{int}(x)\Bigl],
\end{equation}
with $V^{(i)}_{int}(x)=U_{i}\delta(x)$ the band-dependent interface potential controlling the tip-superconductor coupling and $m_{S_{i}}$ the mass of the $i-$th superconducting branch. Similarly, in the normal side of the junction the eigenvalues problem is obtained from Eq.~(\ref{eq:hamiltonian}) making the substitutions: $\Psi_{S_{i}}(x)\rightarrow\Psi_{N}(x)$, $\Delta_{i} \rightarrow 0$, $m_{S_{i}}\rightarrow m_{N}$, $U_{i}\rightarrow 0$.
Once the eigenvalues problem has been solved for each branch, the branch wave functions $\Psi_{N}(x)$ and $\Psi_{S_{i}}(x)$, $i=1,...,~N$, are written in terms of eigenstates of the local Hamiltonians, i.e.:
\begin{eqnarray}
\label{eq:scstates}
&&\Psi_{N}(x)=\frac{(e^{i k_{N} x}+b e^{-i k_{N} x})}{\sqrt{k_{N}}} |e; N\rangle+\frac{a e^{i k_{N} x}}{\sqrt{k_{N}}} |h; N\rangle \nonumber\\
&&\Psi_{S_{i}}(x)=\frac{t^{(i)}_{e} e^{i k_{i} x}}{\sqrt{k_{i}}} |e; S_{i}\rangle +\frac{t^{(i)}_{h}e^{-i k_{i} x}}{\sqrt{k_{i}}} |h; S_{i}\rangle,
\end{eqnarray}
with $a$ and $b$ the Andreev and normal reflection coefficients, while $t^{(i)}_{e/h}$ are transmission coefficients as electron and hole quasiparticles.
The electron and hole spinors of the normal side, $|e; N\rangle=(1,0)^{t}$ and $|h; N\rangle=(0,1)^{t}$, are used to express the ones of the superconducting side
\begin{eqnarray}
&&|e; S_{i}\rangle=u_{i}|e; N\rangle+v_{i}e^{-i\phi_{i}}|h; N\rangle\nonumber\\
&&|h; S_{i}\rangle=v_{i}e^{i\phi_{i}}|e; N\rangle+u_{i}|h; N\rangle,
\end{eqnarray}
while $u_{i}^2=1-v_{i}^2=(1+\sqrt{E^{2}-|\Delta_{i}|^2}/E)/2$ are the usual coherence factors.
In writing Eqs.~(\ref{eq:scstates}), we have taken the wavevectors of the electron-like $|e; N/(S_{i})\rangle$ and hole-like $|h; N/(S_{i})\rangle$ quasiparticles as equal (Andreev approximation), while we fixed $k_{N}=\sqrt{2m_{N}E_{F}/\hbar^{2}}=k_{F}$, $k_{i}=r_{i}k_{F}$, being $r^{2}_{i}=m_{S_{i}}/m_{N}$ the effective mass ratio. In order to avoid unessential technicalities, hereafter we focus on the \textit{metallic limit} of the interface ($V^{(i)}_{int}(x)\rightarrow 0$). The latter assumption does not affect too much the results; indeed, taking different effective masses in the network branches simulates a band-sensitive interface potential which favors tunneling processes into a specific band of the superconducting region. Thus, the low-transparency limit is also included in this description.\\
The particle wave functions must be single-valued at the network node $x=0$, thus implying the following matching conditions:
\begin{equation}
\label{eq:BCs1}
\Psi_{N}(x=0)=\Psi_{S_{1}}(x=0)=...=\Psi_{S_{N}}(x=0).
\end{equation}
The probability current in the branch $N$ ($S_{i}$) is given by $J_{N/S_{i}}(x=0)=\hbar~m^{-1}_{N/S_{i}}\mathrm{Im}[\Psi_{N/S_{i}}^{\dagger}(x)\partial_{x}\hat{\sigma}_{z}\Psi_{N/S_{i}}(x)]_{x=0}$ and thus the associated conservation law is $J_{N}(x=0)=\sum_{i}J_{S_{i}}(x=0)$, where $\hat{\sigma}_{z}$ is the Pauli matrix. Combining Eqs.~(\ref{eq:BCs1}) with the probability current conservation law, we obtain a second set of boundary conditions:
\begin{equation}
\label{eq:BCs2}
\sum_{i}r^{-2}_{i}\partial_{x}\Psi_{S_{i}}(x)\Bigl|_{x=0}-\partial_{x}\Psi_{N}(x)\Bigl|_{x=0}=0.
\end{equation}
Eqs.~(\ref{eq:BCs1})-(\ref{eq:BCs2}) completely define the scattering problem and allow the computation of the Andreev and normal reflection coefficients, namely $a$ and $b$, and $t_{e/h}^{(i)}$. The probability currents $J_{N}$, $J_{S_{i}}$ can be written in terms of the scattering coefficients as $J_{N}=\hbar~m^{-1}_{N}\big(1-|b|^2-|a|^2 \big)$ and $J_{S_{i}}=\hbar~m^{-1}_{S_{i}}\Omega_{i} \big(|t^{(i)}_{e}|^2+|t^{(i)}_{h}|^2 \big)$, with $\Omega_{i}=|u_{i}|^2-|v_{i}|^2$. From the current conservation law it follows that the sum of the Andreev and normal reflection probabilities, $|a|^2$ and $|b|^2$, and of transmission as electron-like or hole-like quasiparticle in the $i-$th band, $r^{-2}_{i}\Omega_{i}|t^{(i)}_{e}|^2$ and $r^{-2}_{i}\Omega_{i}|t^{(i)}_{h}|^2$, add up to one, i.e.
\begin{equation}
|a|^2+|b|^2+\sum_{i} r^{-2}_{i}\Omega_{i}\big(|t^{(i)}_{e}|^2+|t^{(i)}_{h}|^2\big)=1.
\end{equation}
The function $\Omega_{i}(E)$, which controls the band-sensitive tunneling, vanishes when the quasiparticle energy $E<|\Delta_{i}|$, while for $E\gg|\Delta_{i}|$ it asymptotically approaches 1. Since the behavior of $\Omega_{i}(E)$ strongly depends on the considered band, an inhomogeneous distribution of the probability current among the superconducting branches $S_{i}$ is expected. The inhomogeneous distribution of the particles flux is also controlled by different values of band effective mass $m_{S_{i}}$ (within the parabolic band approximation) inside the superconductor. The charge current $I$ flowing through the tip is given by the standard expression \cite{btk82}:
\begin{equation}
I \propto \int dE [f(E-eV)-f(E)][1+|a(E)|^2-|b(E)|^2],
\end{equation}
where $f(E)$ is the Fermi function and $V$ represents the voltage bias across the interface. At zero temperature, the differential conductance $G(V)=dI/dV$ of the system normalized to the normal state conductance $G_N$, takes the following form:
\begin{equation}
\label{eq:diffcond}
\frac{G(V)}{G_{N}}=\frac{1+|a(E=eV)|^2-|b(E=eV)|^2}{1-|b(E \rightarrow\infty)|^2},
\end{equation}
while effects of finite quasiparticle lifetime \cite{dynes78} ($\tau=\hbar/\Gamma$) can be included substituting $E\rightarrow E-i\Gamma$ in Eq.~(\ref{eq:diffcond}).\\
\textit{Results for a two-band model}. We first focus on the relevant case of a two-bands superconductor characterized by $\Delta_{i}=|\Delta_{i}|e^{i\phi_{i}}$, $i=\{1,2\}$, and effective masses $m_{S_{1/2}}$, while we assume that the effective mass $m_{N}$ coincides with the bare electron mass. Without loss of generality, we set $\phi_{1}=0$, $\phi_{2}=\delta$, $|\Delta_{2}|=\gamma \Delta_{1}$ and measure the energy in unit of $\Delta_{1}$ \cite{note-gamma}. The case we are treating describes at least two distinct physical situations: (i) a superconductor with two bands both coupled to the normal side of the junction; (ii) a multiband superconductor in which only two bands are coupled to the normal side of the system. In the former case, detecting $\delta=0$ or $\delta=\pi$ reflects the internal pairing symmetry $s_{++}$  or  $s_{\pm}$ of the system, while, in the latter case, the detection of a phase difference $\delta=0$ or $\pi$ between the two order parameters could be the effect of the (contact-dependent) coupling among tip states and a small subset of the superconducting bands \cite{nota_directional_tunneling}.\\
Using Eqs.~(5) and (6), we derived the Andreev and normal reflection coefficients as reported below
\begin{widetext}
\begin{eqnarray}
\label{eq:coeff}
&&a=\frac{4r_{1}r_{2}(r_{1}w_{1} u_{2}v_{2}+e^{i\delta}r_{2}w_{2} u_{1}v_{1})}{e^{i\delta}[w_{1}w_{2}(r^{2}_{1}r^{2}_{2}+r^{2}_{1}+r^{2}_{2})+2r_{1}r_{2}(1+r_{1}w_{1}+r_{2}w_{2})]-8r_{1}r_{2}u_{1}v_{1}u_{2}v_{2}}\\\nonumber
&&b=\frac{8r_{1}r_{2}u_{1}v_{1}u_{2}v_{2}+e^{i\delta}[w_{1}w_{2}(r^{2}_{1}r^{2}_{2}-r^{2}_{1}-r^{2}_{2})-2r_{1}r_{2}]}{e^{i\delta}[w_{1}w_{2}(r^{2}_{1}r^{2}_{2}+r^{2}_{1}+r^{2}_{2})+2r_{1}r_{2}(1+r_{1}w_{1}+r_{2}w_{2})]-8r_{1}r_{2}u_{1}v_{1}u_{2}v_{2}},
\end{eqnarray}
\end{widetext}
where $w_{i}=u^{2}_{i}-v^{2}_{i}$. Eqs.~(\ref{eq:coeff}) represent the main result of this work and include mass mismatch and interference effects between the two order parameters. The inclusion of appropriate effective masses for the superconducting bands is a crucial ingredient to account for the partitioning of the incoming current among the superconducting channels. For $E\rightarrow \infty$, the Andreev reflection coefficient goes to zero, while $b(E\rightarrow \infty)=(r_{1}r_{2}-r_{1}-r_{2})/(r_{1}r_{2}+r_{1}+r_{2})$. In the absence of mass mismatch ($r_{1}=r_{2}=1$), the transmission probability of the junction in the normal state, $1-|b(E\rightarrow \infty)|^2$, takes the value $8/9$, indicating that the bottleneck effect induced by the Y-junction enhances the reflection probability compared to a single channel geometry. Within the standard (single channel) BTK picture a transmission probability $\mathcal{T}$ of $8/9$ would be obtained by considering a delta-like potential characterized by a BTK parameter $Z_{BTK}=\sqrt{2}/4 \approx 0.35$ (notice that $\mathcal{T}=(1+Z^{2}_{BTK})^{-1}=8/9$). In order to study the interference effects between the two order parameters, we consider the zero energy limit of Eqs.~(\ref{eq:coeff}):
\begin{eqnarray}
\label{eq:coeff0}
&&a(E\rightarrow 0)=-\frac{2ir_{1}r_{2}(r_{1}+e^{i\delta}r_{2})}{2r_{1}r_{2}+e^{i\delta}(r^{2}_{1}r^{2}_{2}+r^{2}_{1}+r^{2}_{2})}\\\nonumber
&&b(E\rightarrow 0)=\frac{e^{i\delta}(r^{2}_{1}r^{2}_{2}-r^{2}_{1}-r^{2}_{2})-2r_{1}r_{2}}{2r_{1}r_{2}+e^{i\delta}(r^{2}_{1}r^{2}_{2}+r^{2}_{1}+r^{2}_{2})}.
\end{eqnarray}
For the $s_{\pm}$ pairing ($\delta=\pi$), Eqs.~(\ref{eq:coeff0}) show that the Andreev reflection is strongly suppressed and eventually vanishes for $r_{1}=r_{2}$. In the latter case, the suppression of the Andreev mechanism combined with the absence of available quasiparticle states below the gap in the superconducting side of the junction produces the total (normal) reflection of the incident particle ($|b|^2=1$). Under this condition the zero-bias conductance is zero (Fig.~2~(a)), while the conduction is restored when $r_{1}\neq r_{2}$ (Fig.~2~(c)).
\begin{figure}[!]
\includegraphics[clip,scale=0.5]{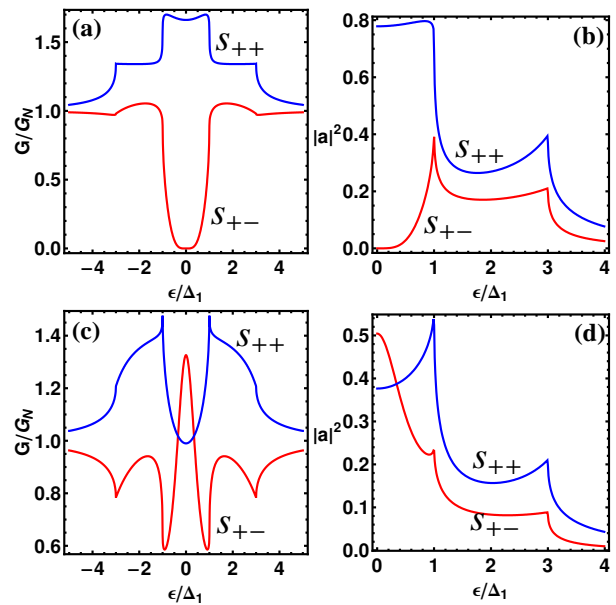}\\
\caption{(Color online) Normalized differential conductance $G/G_{N}$ \textit{vs} $\epsilon/\Delta_{1}$ (a) and the Andreev reflection probability $|a|^2$ (b) obtained by fixing $\gamma=3$, $r_{1}=r_{2}=1.2$. The lower panels (c), (d) are obtained assuming $\gamma=3$, $r_{1}=0.6$, $r_{2}=0.8$. Notice the completely destructive interference effect of the Andreev reflection in the $s_{\pm}$ case (a)-(b), and the restoring of zero-bias conduction as the condition $r_{1}\neq r_{2}$ is met (c)-(d).}
\label{fig:fig2}
\end{figure}
In real systems, the complete suppression of the zero-bias Andreev reflection is hindered by the different values of effective mass which usually characterize the superconducting bands. For the $s_{++}$ pairing ($\delta=0$), the presence of two bands only produces a reduction of the zero-bias Andreev reflection probability (Fig.~\ref{fig:fig2} (b)) and a corresponding enhancement of the normal reflection probability. All these features are illustrated in Fig.~\ref{fig:fig2} where the differential conductance $G/G_{N}$ and the Andreev reflection probability $|a|^2$ are reported by fixing $\gamma=3$, $r_{1}=r_{2}=1.2$ in the upper panels (a)-(b), and $\gamma=3$, $r_{1}=0.6$, $r_{2}=0.8$ in the lower panels (c)-(d). The mechanism leading to the formation of a zero-bias peak starting from a completely suppressed conductance is explored in Fig.~\ref{fig:fig3} where $G/G_{N}$ \textit{vs} $\epsilon/\Delta_{1}$ is given for $s_{\pm}$ (a) and $s_{++}$ (b) symmetry. For the $s_{\pm}$ case, taking $\gamma=2$ and $r_{1}=\sqrt{0.75}$, the zero bias conduction is restored as the mass ratio $r_{2}$ is increased. For the $s_{++}$ case a double peak structure, less sensitive to the $r_{2}$ change, is always present.
Differential conductance curves similar to the one reported in Figs.~(\ref{fig:fig2})-(\ref{fig:fig3}) have already been experimentally found in Refs. \cite{Naidyuk2014,daghero2012,fetese-multigap} giving evidence of
 both $s_{++}$ and $s_{\pm}$ pairing symmetry. This is only an apparent ambiguity since it has been shown that, under appropriate circumstances, the $s_{\pm}$ symmetry can be converted into the $s_{++}$ pairing in the vicinity of a reflecting surface \cite{bobkov}.\\
\begin{figure}[!t]
\includegraphics[clip,scale=0.55]{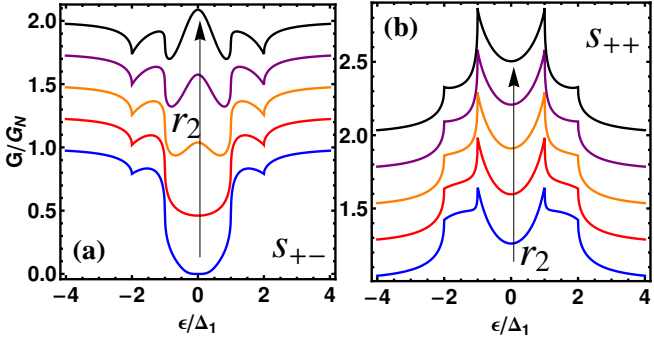}\\
\caption{(Color online) Normalized differential conductance $G/G_{N}$ \textit{vs} $\epsilon/\Delta_{1}$ for an N/$s_{\pm}$ interface (a) and N/$s_{++}$ interface (b). The model parameters have been fixed as follows: $\gamma=2$, $r_{1}=\sqrt{0.75}$, while $r_{2}$ takes values $\sqrt{0.75}$ (bottom curve), $1$, $\sqrt{1.25}$, $\sqrt{1.5}$, $\sqrt{1.8}$ (top curve). An offset of $(n-1)/4$ has been added to the nth curve. In the N/$s_{\pm}$ case, the reactivation of the Andreev reflection induced by an $r_{2}\neq r_{1}$ gradually induces a zero-bias peak surrounded by two dips located at energy $\epsilon \approx \Delta_{1}$ and  $\epsilon=\Delta_{2}$, respectively.}
\label{fig:fig3}
\end{figure}
The low-transparency regime of the system is examined in Fig.~\ref{fig:fig4} where the differential conductance of the N/$s_{\pm}$ (a) and N/$s_{++}$ (b) junction is reported by fixing $\gamma=3$, $r_{1}=0.7$ and letting $r_{2}$ varying from $0.15$ to $0.65$ as in the figure caption. While the differential conductance of the N/$s_{++}$ case presents a single minimum for $\epsilon<\Delta_{1}$, interesting subgap structures are present in the N/$s_{\pm}$ case. In the latter case (Fig.~\ref{fig:fig4} (a)), the increasing of $r_{1}$ produces the coalescence of two conductance peaks, which form a single zero bias structure for $r_{1} \geq 0.5$. A further increasing of $r_{1}$ first produces a lowering of the peak amplitude and then, for $r_{1}\approx r_{2}$, a completely suppressed conductance. These subgap structures are associated to Andreev bound states which manifest themselves as complex poles of the Andreev reflection coefficient (see Eqs.~(\ref{eq:coeff})). The pole energies are solution of the equation ($\xi=\epsilon/\Delta_{1}$)
\begin{eqnarray}
\label{eq:poles}
&&\sqrt{(1-\xi^2)(\gamma^2-\xi^2)}\frac{r^{2}_{1}r^{2}_{2}+r^{2}_{1}+r^{2}_{2}}{2 r_{1}r_{2}}-\xi^2+\\\nonumber
&&-ir_{1}\xi \sqrt{1-\xi^2}-ir_{2}\xi \sqrt{\gamma^2-\xi^2}+\gamma e^{-i\delta}=0
\end{eqnarray}
coming from the divergence condition of the scattering coefficients ($a$ or $b$). We numerically solved Eq.~(\ref{eq:poles}) setting the model parameters as in Fig.~\ref{fig:fig4} (a) ($s_{\pm}$ case). The solution presents poles of the form $\xi_{\pm}=\pm \xi_{R}-i\xi_{I}$ having opposite real part and coincident imaginary part. According to Ref.~\cite{golubovprl2009}, no bound states exist at energies between the two gaps. In Fig.~\ref{fig:fig4} (c) we show the real part $\xi_{R}=\epsilon_{R}/\Delta_{1}$ (dot symbols) and the imaginary part $\xi_{I}=\epsilon_{I}/\Delta_{1}$ (vertical bars) of $\xi_{+}$ evaluated at different values of $r_{1}$. By increasing $r_{1}$, $\xi_{R}$ goes to zero and, correspondingly, $\xi_{I}$ increases. When the condition $\xi_{I} \gg \xi_{R}\approx 0$ is reached, the two poles
 $\xi_{+}$ and $\xi_{-}$ cause a single zero-bias peak in the conductance (Fig.~\ref{fig:fig4} (a)), despite their real parts remain distinct. The above arguments correlate the pole structure of the scattering coefficients with the subgap features peculiar of the $s_{\pm}$ symmetry.\\
\begin{figure}
\includegraphics[clip,scale=0.52]{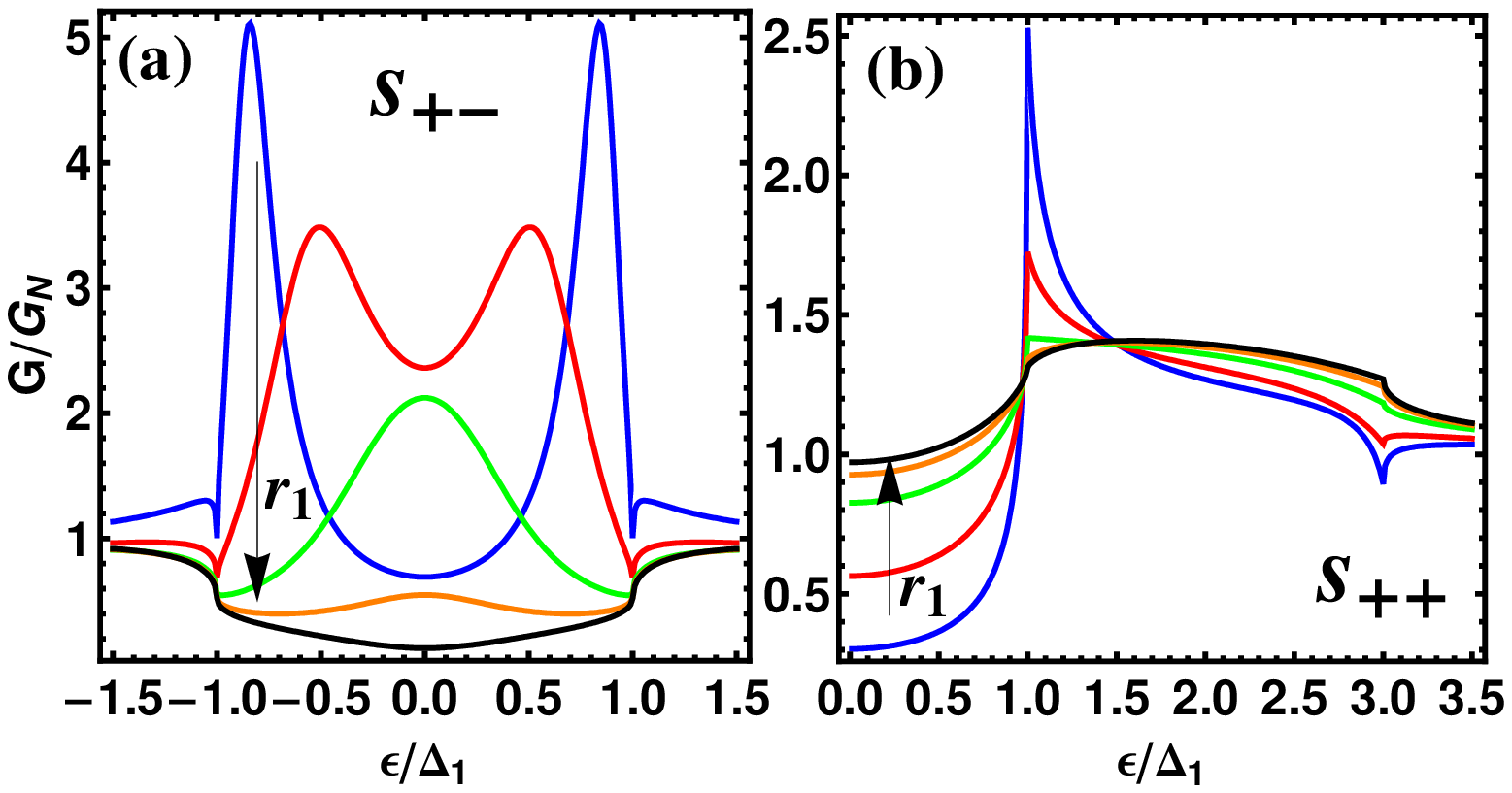}\\
\includegraphics[scale=0.72]{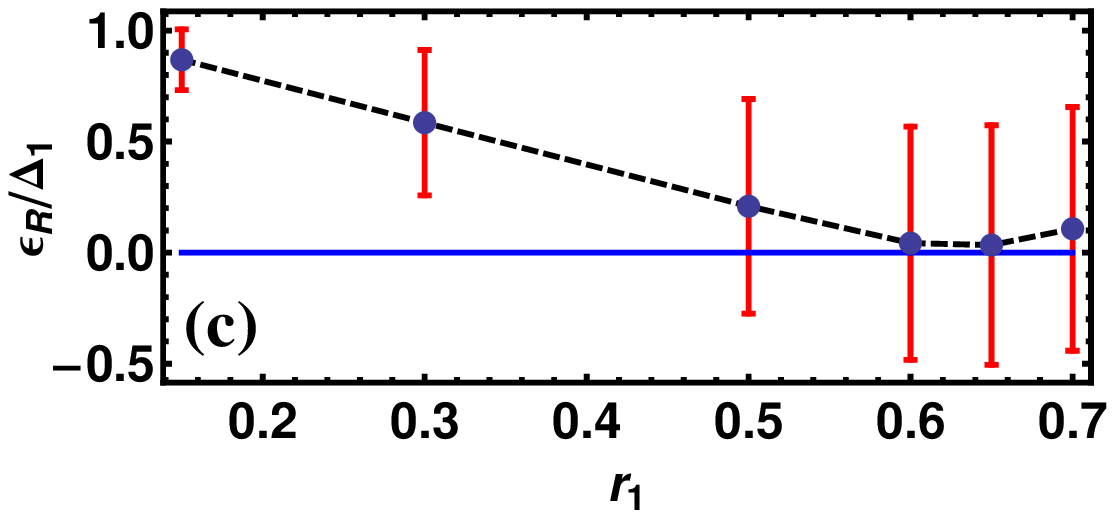}
\caption{(Color online) Normalized differential conductance $G/G_{N}$ \textit{vs} $\epsilon/\Delta_{1}$ for an N/$s_{\pm}$ interface (a) and N/$s_{++}$ interface (b). The model parameters have been fixed as follows: $\gamma=3$, $r_{2}=0.7$, while $r_{1}$ takes values 0.15 (arrow tail), 0.3, 0.5, 0.6, 0.65 (arrow head). For the N/$s_{\pm}$ interface with $r_{1}=0.5$, a zero-bias peak is formed by the fusion of two conductance peaks, the latter structure being absent in the $s_{++}$ case (b). (c) $\epsilon_{R}/\Delta_{1}$ \textit{vs} $r_{1}$ fixing the model parameters as in panel (a), $\pm \epsilon_{R}-i\epsilon_{I}$ being the poles of the Andreev reflection coefficient. The vertical bars indicate the imaginary part $\epsilon_{I}/\Delta_{1}$ of the pole, while the dashed line is a guide for the eyes.}
\label{fig:fig4}
\end{figure}
In conclusion, we provided a minimal model of the transport properties of a normal-metal/multiband superconductor interface based on the quantum waveguides theory. For a two-band superconductor ($s_{\pm/++}$), an analytic expression of the conductance is derived, showing, for the $s_{\pm}$ pairing, destructive interference between the order parameters. The mass mismatch among the superconducting bands is taken into account and determines, together with the gap ratio $\Delta_{2}/\Delta_{1}$, how the incident particles flux is partitioned among different conducting channels. The theory captures the genuine multiband nature of the problem and describes effects which do not simply derive from an incoherent sum of distinct single-band tunneling probabilities.\\
\textit{Acknowledgements}. The authors acknowledge F. Giubileo for useful discussions on tunneling spectra analysis.


\begin{thebibliography}{99}
\bibitem{note-ballistic} The ballistic-contact assumption does not limit our analysis since diffusive contacts can be modeled using a ballistic theory with reduced interface transparency.
\bibitem{note-gamma} In the following we assume a gap ratio $\gamma=2$ or $\gamma=3$ which is an appropriate choice for the materials of interest.
\bibitem{nota_directional_tunneling} As also found in the MgB$_{2}$ case, the possibility to observe a certain number of gaps in point-contact spectra could depend on the tunneling direction, barrier type and amount of impurities.
\end{thebibliography}
\end{document}